\documentclass[twocolumn]{aastex63}

\turnoffeditone
\turnoffedittwo  

\accepted{October 21, 2020}
\submitjournal{ApJ}
\shorttitle{Fan Beam Model in Population Synthesis of Pulsars}
\shortauthors{Huang \& Wang}

\begin{document}

	\title{Role of Fan Beam Model in Population Synthesis of Isolated Radio Pulsars}
	\correspondingauthor{H. G. Wang}
	\email{wjhuang@e.gzhu.edu.cn, hgwang@gzhu.edu.cn}

	\author{W. J. Huang}
	\affiliation{School of Physics and Electronic Engineering, Guangzhou University, 510006 Guangzhou, China}

	\author[0000-0002-2044-5184]{H. G. Wang}
	\affiliation{School of Physics and Electronic Engineering, Guangzhou University, 510006 Guangzhou, China}
    \affiliation{Xinjiang Astronomical Observatories, Chinese Academy of Sciences, Urumqi 830011, China}

	\begin{abstract}
		On the basis of the \textsf{P{\scriptsize SR}P{\scriptsize OP}P{\scriptsize Y}} software package, we developed an evolution model of population synthesis for isolated radio pulsars by incorporating the fan beam model, an alternative radio emission beam model to the widely used conal beam model. With proper choice of related models and parameters, we succesfully reproduced the observational distributions of Galactic longitude ($l$) and latitude ($b$), spin period, period derivative, dispersion measure and 1.4-GHz \edit1{flux density} for the 1214 isolated pulsars discovered by the Parkes multibeam and \edit1{Swinburne} pulsar surveys. The number of underlying population of radio-loud pulsars is found to be $2.27\times10^6$, which is larger than the underlying population of radio-loud pulsars simulated with conal beam model. The model is used to estimate the number of isolated radio pulsars detected by the Galactic-plane pulsar survey with Five hundred-metre Aperture Spherical Telescope (FAST). Approximately \edit1{2700} and \edit1{240} new isolated pulsars are expected to be discovered in the inner galactic plane ($20^{\circ} < l < 90^{\circ}$, $|b| < 10^{\circ}$) and the outer galactic plane ($150^{\circ} < l < 210^{\circ}$, $|b| < 10^{\circ}$), respectively. These numbers are smaller than those estimated by the evolution models with conal beam and the snapshot models. 

	\end{abstract}

	\keywords{radio pulsars, population synthesis, fan beam model, FAST survey}
	
	\section{Introduction} \label{sec:introduction}
       
		In order to understand the underlying distribution of parameters of Galactic pulsars, it is general to perform Monte Carlo (MC) simulations of pulsar populations to generate artificial pulsars that satisfy the criteria of detection of exsisting surveys, and then infer the distributions by comparing the simulated pulsars and those detected in actual surveys. This method of population synthesis has been widely used to study the initial parameters and their evolution for radio pulsars (e.g. \citealt{2002ApJ...568..289A}; \citealt[][hereafter FK06]{2006ApJ...643..332F}; \citealt[][hereafter RL10]{2010MNRAS.404.1081R}; \citealt{2020MNRAS.492.4043C}) and high-energy pulsars \citep[e.g.][]{2004ApJ...604..775G,2013ApJ...776...61P,2018MNRAS.481.3966B}. It has also been used to estimate the size of pulsar population or subpopulation in the Galaxy \citep[e.g.][]{1990ApJ...352..222N,2006MNRAS.372..777L,2014ApJ...787..137S} and the \edit1{nearby} galaxies (\citealt[][hereafter S09]{2009A&A...505..919S}; \citealt{2010MNRAS.406L..80R}), and to predict yields for pulsar surveys with radio telescopes like FAST and SKA \citep[S09;][]{2020MNRAS.492.4043C}, or with the Fermi gamma-ray telescope \citep[e.g.][]{2018ApJ...863..199G}.

        In the population synthesis for radio pulsars, whether a synthesic pulsar can be detected by a given survey is directly affected by four parameters of the pulsar: spin period ($P$), pulse width ($W$), radio luminosity ($L$) and distance ($d$). These parameters, together with other parameters of the teleslcope and survey, are used to estimate the signal to noise ratio ($S/N$) of the radio pulse. Only when the $S/N$ is higher than a certain level (e.g. 9) can the pulsar be regarded as being detected. The parameters $W$ and $L$ are related to the physics of radio emission, among which $W$ depends on specific emission beam model. The conal beam model has been widely used in population synthesis, in which the cross section of radio emission beam is usually \edit1{assumed} to be circular. An empirical relationship $\rho = \rho_{0}P^{-1/2}$ \citep{1990ApJ...352..247R,1998ApJ...501..270K} is often used to generate a beam radius for a synthetic pulsar, and then $W$ can be calculated with the parameters $\rho$, the inclination angle between the spin and magnetic axes ($\chi$) and the impact angle between the line of sight (LOS) and the magnetic axis ($\beta$). The radio luminosity is usually assumed to have a power-law dependence on $P$ and $\dot{P}$ (period derivative), i.e. $L = L_{0}P^{q_1}\dot{P}^{q_2}$, of which the indices are constrained to be different values by a number of studies, mostly within the range $-1.6<q_1<-0.9$ and $0.3<q_2<0.6$ \edit1{\citep[cf.][]{2014MNRAS.443.1891G}}.

		The double-conal beam model \citep{1983ApJ...274..333R,1993ApJ...405..285R}, the most popular form of conal beam model, which suggests that the radio beam consists of two nested cones and a core, has the advantage in explaining a variety of pulse morphology in terms of different LOS trajectories across the beam. The fan beam model, an alternative radio emission model originally proposed by \citet{1987ApJ...322..822M}, has been developed in the past decade (\citealt{2010MNRAS.401.1781D}; \citealt{2012MNRAS.420.3403D,2013MNRAS.434.3061D}; \citealt[][hereafter W14]{2014ApJ...789...73W}; \citealt{2015MNRAS.446.2505D}; \citealt{2019MNRAS.489..310O}). This kind of model assumes that relativistic particles in the pulsar magnetosphere may produce a radially elongated fan beam. Unlike the conal beam, the fan beam may be underfilled in longitude. They can explain some phenomena that are hard to explain by the conal beam models.
		
		Several pieces of evidence or indications of the fan beam model have been presented in the literature, mainly including (a) the observed fan-like radio beams of the precessional pulsar PSR J1906$+$0746 \citep{2013IAUS..291..199D,2019Sci...365.1013D}, (b) the  possitive correlation between the observed pulse width and the absolute value of the impact angle as predicted by the fan beam model (W14, note that the conal beam model predicts a negative correlation), (c) the successful reproduction of observational phenomena such as the bifurcated emission component, the diversity in the pulse morphology and the radius-to-frequency mapping \citep{2010MNRAS.401.1781D,2012MNRAS.420.3403D,2015MNRAS.446.2505D}, (d) the more natural interpretation of some phenomena with the fan beam model than with the conal beam model, e.g. the ratio of pulse peak separation \citep{2015MNRAS.454.2216D}, the beam properties of radio pulsars with interpulse emission \citep{2019MNRAS.490.4565J} and the single pulse frequency evolution of PSR J1136+1551 \citep{2019MNRAS.489..310O}.
		
		The fan beam model by W14 gives a different relationship for $W$ from the conal beam model, which depends on $\chi$, $\beta$ and the longitudinal extent of the flux tube active for radio emission. The model also gives a semiempirical relationship between the emission intensity and three parameters: $P$, $\dot{P}$ and the angular distance of a given point in the beam to the magnetic pole (hereafter called radial distance). The general feature of the intensity ditribution in the beam is that the intensity decreases with the increasing radial distance in the outer part of the beam (roughly beyond 2 times of the opening angle of the polar cap). Comparing with a simple conal beam model, in which the intensity is usually assumed to be uniform, the fan beam model suggests a radially dependent intensity pattern, therefore the luminosity function would be different from that of the conal beam model. In this sense, it is necessary to explore the population synthesis by incorporating in the fan beam model, in order to see how this model affects our understanding of pulsar population.  
	
		There are mainly two different techniques for population studies: ``snapshot'' and ``evolution'' \citep{2019BAAS...51c.261L}. In the snapshot approach, pulsars are generated from a group of optimal distribution functions derived from the observed sample, which are typically of the parameters $P$, $L$, Galactocentric radius ($R$) and Galactic height ($z$) \citep[hereafter L06]{2006MNRAS.372..777L}. While in the evolution approach, the model pulsars are allowed to evolve over time from a set of initial distributions, such as birth velocity, location in the Galaxy and initial spin period (cf. FK06).
		
		The \textsf{P{\scriptsize SR}P{\scriptsize OP}P{\scriptsize Y}}\footnote{https://github.com/samb8s/PsrPopPy} \citep[][hereafter B14]{2014MNRAS.439.2893B}, rewritten from \textsf{P{\scriptsize SRPOP}}\footnote{http://psrpop.sourceforge.net/} package (cf. L06) in Python, is one of a few open-source software packages for pulsar population synthesis. With many callable programs for specific distributions or optimal models (including the conal beam model), it can carry out both the snapshot and evolution approaches. It has been used in a couple of population studies, e.g. predictions for the future SKA pulsar survey \citep{2015aska.confE..40K} and continuum surveys using variance imagining \citep{2017MNRAS.472.1458D}.
		
	    The \textsf{P{\scriptsize SR}P{\scriptsize OP}P{\scriptsize Y}} software is convenient for users to add new models. In this paper, we aim to develope a population synthesis code on the basis of \textsf{P{\scriptsize SR}P{\scriptsize OP}P{\scriptsize Y}} by adding a module for the fan beam model. Since the relationships given by W14 are only for normal pulsars, this work is limited to model the Galactic population for isolated normal pulsar. The updated code\footnote{https://github.com/wjhgakki/PsrPopPy\_WJH} is then applied to estimate the population size of radio-loud isolated pulsars and the expected number of discovery by the FAST pulsar survey. The result will be compared with a former work on the yields of FAST survey by S09, which is on the basis of the conal beam model. MC simulation procedures and related models are described in Section \ref{sec:method}. Simulations on pulsar population are introduced in Section \ref{sec:Pop_Syn}. The prediction on the yields of FAST pulsar survey is presented in Section \ref{sec:FAST_dectection}.  Conclusions and discussions are made in Section \ref{sec:conclusions}.
		
	\section{Models in the simulation} \label{sec:method}
	    
	    In this paper, the model pulsar populations are generated by using the \textsf{\small{EVOLVE}} code of the \textsf{P{\scriptsize SR}P{\scriptsize OP}P{\scriptsize Y}} package, corresponding to the evolution approach. In order to compare the fan beam and conal beam models, we generate evolution models A and B (hereafter simply called model A for the conal beam and model B for the fan beam). Various optimal models from previous works (e.g. FK06; S09; RL10) are adopted for model A, while in model B the original relations of pulse width and radio luminosity are replaced with those in the fan beam model. Note that all frequency-dependent models are based on the analysis at 1.4 GHz, the frequency of the pulsar surveys used in this paper, i.e. the Parkes multibeam pulsar survey (PKSMB) near the Galactic plane \citep{1996MNRAS.279.1235M} and the two Swinburne pulsar surveys (PKSSW) at higher Galactic latitudes \citep{2001MNRAS.326..358E, 2009ApJ...699.2009J}. 
	    	
	  	In the following we first describe the models of some parameters that have difference in models A and B, then we outline the other models commonly used in the two synthetical models. 
	  	    
    \subsection{Modelling Pulse Widths}
    
    \subsubsection{Model A}
    
        In modle A, the method of modelling the pulse width is the same as that used by S09, on the basis of a relationship between $W$ and two geometric parameters, $\chi$ and $\beta$, namely
        \begin{equation}{}
            \sin^{2}(\frac{W}{4}) = \frac{\sin^{2}(\frac{\rho}{2})-\sin^{2}(\frac{\beta}{2})}{\sin \chi  \sin(\chi+\beta)} {},
            \label{eq:conal_beam_width}
        \end{equation}
        where $\rho$ is the radius of the circular beam. $\rho$, $\chi$ and $\beta$ are all random parameters for a population, and their distributions must be assigned first. 
        
        Following the empirical relationship between $\rho$ and $P$ found by \citet{1998ApJ...501..270K}, 
        \begin{equation}
            \rho_{{\rm model}} = \left \{ \begin{array}{ll}
            5.4^{\circ} P^{-1/2} & P>30 \ {\rm ms} \\
            31.2^{\circ} & {\rm otherwise} 
            \end{array} \right.,
        \end{equation}
        an initial value of beam radius is generated. Then the value is dithered by a random value $p$ to model the dispersion of $\rho$, which reads
        \begin{equation}
            \rho = 10^{{\rm log}_{10}(\rho_{{\rm model}})+p}{},
        \end{equation}
        where $p$ is drawn from a uniform distribution between $-0.15$ and 0.15. 
        
        Since the LOS must sweep across the beam so that the synthetic pulsar could be detected, $\beta$ can be simply chosen from a flat distribution as $-\rho \leq \beta \leq \rho$ for radio-loud pulsars that are potentially be detected. 
        
        As for $\chi$, a number of works suggested that the magnetic axis tends to be aligned with the spin axis, i.e. $\chi$ decreases to 0 eventually (e.g. \citealt[][hereafter TM98]{1998MNRAS.298..625T}; \citealt{2008MNRAS.387.1755W}; RL10; \citealt{2017ApJ...837..117T}). However, some works proposed that the inclination angle may evolve to $90^{\circ}$ \citep[e.g.][]{2013Sci...342..598L,2017MNRAS.466.2325A}. Since there is no concensus yet on the evolution of inclination angle, in this paper, following S09, we consider a simple geometric case that $\chi$ is chosen from a random distribution, i.e. 
        \begin{equation}
            \chi = \arccos u, 	
            \label{eq:chi}
        \end{equation}
        where $u$ is drawn from a flat distribution between 0 and 1. This method of generating $\chi$ is also used in model B.
        
    \subsubsection{Model B}
        
        In model B, the treatment of pulse width is different from model A. W14 has indicated that the way to calcuate pulse width is different for the inner and outer parts of a fan beam. In the outer beam, the pulse width widens as $|\beta|$ increases, following the relationship
        \begin{equation}
            \cos(\frac{W}{2}+C) = \frac{\sin \chi}{\tan (\chi+\beta)(\cos^{2}\chi+\tan^{-2}\varphi)^{1/2}},
            \label{eq:fan_beam_width}
        \end{equation}
        where $C=\arctan(\sec \chi/\tan \varphi)$ and $\varphi$ is the half azimuthal width of flux tube.
        
        This relationship is a consequence of the divergence nature of the dipolar flux tube. However, the relationship for the inner part is quite uncertain because  subbeams from multiple flux tubes (if exist) may be bright enough to be observed, hence the equation may not be so simple as Eq. (\ref{eq:fan_beam_width}). The boundary dividing the inner and outer parts is close to the polar cap. By modeling the correlation of the observed pulse widths and impact angles for a sample of 64 pulsars, the boundary was found to be $\rho\simeq2\rho_{\rm pc}$, where the opening angle of the polar cap satisfies $\rho_{{\rm pc}} \simeq 3/2(R/R_{{\rm c}})^{1/2}$, with $R$ the stellar radius (assumed to be 10 km) and $R_{{\rm c}}$ the light cylinder radius ($R_{{\rm c}} = cP/2 \pi$, where $c$ is the light speed). For simplicity, following W14, we fix the beam radius as $2\rho_{pc}$ when $|\beta| \leq 2\rho_{pc}$ (in the inner beam). Then we could compute the pulse width approximately with Eq. (\ref{eq:conal_beam_width}) by substituting $\rho=2\rho_{{\rm pc}}$ and $\beta=0$. 
        
        In the fan beam model, the half azimuthal width of the flux tube, $\varphi$, is an intrinsic parameter to determine the pulse width. In our simulation $\varphi$ is chosen from a uniformly distribution in [$\varphi_{\rm min}$, $\varphi_{\rm max}$]. According to the simulation by W14, the range is required to be [20$^{\circ}$, 80$^{\circ}$] to match the observed $W$ and $\beta$ for a sample of 64 puslars. In our simulation different choices of the range are also tried, as discussed in the next section.   
        
        The impact angle has to be determined in a different way in model B as well, because there is no abrupt radial boundary of the beam as suggested by W14. It is then generated from the distributions of $\zeta$ and $\chi$ via the relationship $\beta=\zeta-\chi$, where $\zeta$ is the viewing angle between the LOS and the spin axis. Assuming that the projections of the spin axes of pulsars are uniformly distributed in the celestial sphere, the probability density function of $\zeta$ should be
        \begin{equation}
            p(\zeta) = \int_{0}^{2\pi} \frac{1}{4\pi} \sin \zeta d \Phi =\frac{1}{2} \sin \zeta,
        \end{equation}
        where $\Phi$ is the pulse longitude. 
        
        It should be noted that the asumption of \edit1{no} radial boundary may be too simple. It is likely that the radio emission is ineffective if the emission altitude is too high. We set a limit of $|\beta|\leq90^{\circ}$ artificially regarding this point. Equivalent to say, no emission can be observed from a pole when $\beta>90^{\circ}$. However, in the opposite pole, the impact angle is $\beta-180^{\circ}$, of which the absolute value is smaller than $90^\circ$ so that the emission is still observable. Therefore, all the $\beta$ values greater than $90^{\circ}$ are converted to $\beta-180^{\circ}$. 
        
        Note that Eq. (\ref{eq:fan_beam_width}) would be only a rough approximation to the actual pulse width when $|\beta|$ is large, e.g. $|\beta|>40^\circ$ (the corresponding emission height is greater than 20\% of the light cylinder radius), because the aberration, retardation and magnetic field sweep-back effects will become important to affect the pulse width considerably \citep[e.g.][]{2005ApJ...628..923G}. However, as the intensity attenuates dramatically with increasing $|\beta|$, the over-simplification of pulse width at large $|\beta|$ may not cause serious problem. In fact, our simulations do show that most synthetic pulsars detected by the surveys have $|\beta|$ values smaller than 20$^\circ$ (see Section \ref{sec:Pop_Syn}).
      
    \subsection{Beaming Fraction} 
    
        Since the conal beam is only beamed to a fraction of the full sky, a beaming factor is needed to account for the probability that a pulsar beams towards the Earth. The default model in \textsf{\small{EVOLVE}} of beaming fraction is an empirical period-dependent relationship found by TM98, 
        \begin{equation}
            f(P) = 0.09\left[\rm{log(\frac{P}{1s})}-1\right]^{2} + 0.03.
        \end{equation}
        As per RL10, to determine whether a pulsar is beaming towards us or not, the beaming fraction is calculated for a given $P$ and then compared with a random value drawn from a uniform distribution between 0 and 1. Only those pulsars with beaming fraction larger than the random value can be observed by us. 
    
        It should be noted that the above relation is based on the assumption that the beam shape is circular, it is not applicable to the fan beam model. In model B, since we have assumed that the fan beam may extend to $|\beta|=90^\circ$, our LOS would sweep across at least one emission beam from either one pole or the other pole. Therefore $f$ should be set as 1 for model B. 
        
    \subsection{Luminosity Distributions}
        In \textsf{P{\scriptsize SR}P{\scriptsize OP}P{\scriptsize Y}}, B14 provided an option of luminosity distribution obtained by FK06
    	\begin{equation}
        	\edit1{{\rm log}}L = \edit1{{\rm log}}(L_{0} P^{-1.5}\dot{P}^{0.5}_{15})+{\rm L_{corr}},
        	\label{eq:FK06_Lum}
    	\end{equation}
        where $L_{0}$ equals 0.18 mJy kpc$^{2}$, $P$ is in s, $\dot{P}$ is in 10$^{-15}$ s/s, and $L_{{\rm corr}}$ is a dither factor generated from a normal distribution centred on 0 with a standard deviation 0.8, which may partially reflect the errors in the distance measurements of actual pulsars. This method is adopted in model A.
	
        Note that the conal beam model does not provide a picture about the distribution of intensity within the emission cone, thus the luminosity is independent to the beam parameters. On the contrary, the fan beam model propsed by W14 predicts a clear limb-darkening relationship between the peak intensity at 1.4 GHz and the radial distance $\rho_{{\rm peak}}$ in the outer beam ($|\beta|>2\rho_{\rm {pc}}$), i.e. 
	    \begin{equation}
    	    I_{\rm{outer, peak}}^{\rm{1400MHz}} = {\kappa_{0}} P^{q_{0}-4}\dot{P}_{-15} \rho^{2q_{0}-6}_{\rm{peak}},
    	    \label{eq:fan_I}
	    \end{equation}
        where $I_{\rm{outer, peak}}^{\rm{1400MHz}}$ is in unit of ${\rm erg/s/MHz^{-1}/sr}$, $P$ is in unit of second, $\rho_{{\rm peak}}$, the radial distance between the magnetic pole and the emission direction accounting for the pulse peak, is in unit of degree, and $\dot{P}_{-15} = \dot{P}/(10^{-15} {\rm \ s/s})$, respectively. The best-fit values $\kappa_{0} = 10^{25.7\pm1.5}$ and $q_{0} = 1.75 \pm 1.5$ at the 95\% confidence level were obtained statistically in terms of a sample of 64 pulsars by W14. With the best-fit value of $q_{0}$, the intensity attenuates with increasing $\rho_{\rm peak}$ following the law $I_{\rm{outer, peak}}^{\rm{1400MHz}}\propto\rho_{\rm peak}^{-2.5}$ in the outer beam.
		
        We next derive the equation of luminosity with the above relationship. \edit1{Noting that $\rho_{{\rm peak}}$ depends on the pulse profile shape, we simply take $\rho_{{\rm peak}} = |\beta|$ to avoid the complexity of profile shapes in the simulations.} The observed mean flux density reads $F_{{\rm mean}}=\delta F_{{\rm peak}}$, where $\delta\simeq W/P$ is the pulse duty cycle and $F_{{\rm peak}}$ is the peak flux density. Since $F_{{\rm peak}}$ can be converted by $I^{1400 {\rm MHz}}_{{\rm outer, peak}}/d^{2}$ for a given pulsar distance $d$, then the pseudo luminosity in the outer beam should be 
	    \begin{equation}
    	    {\rm log} L = {\rm log}(\kappa \delta P^{q-4}\dot{P}_{-15} |\beta|^{2q-6}) + L_{{\rm corr}},
    	    \label{eq:fan_Lum}
	   \end{equation}
	    where $L$ is in unit of mJy kpc$^2$ \edit1{and $\kappa=10^{2.75}$ is converted from $\kappa_0$}. As per FK06, we also set a dither factor $L_{{\rm corr}}$, which is chosen from a Gaussian distribution with the mean value 0 and the standard deviation 0.75 estimated by the uncertainty of $\kappa_{0}$ in Eq. (\ref{eq:fan_I}). However, in our simulation, $q_0=1.75$ was found not be able to reproduce the observed flux distribution well, hence $q$ is set as a free parameter. For the inner beam ($|\beta| \leq 2\rho_{pc}$), the luminosity is assumed to be a constant calculated with Eq. (\ref{eq:fan_Lum}) by taking $|\beta|=2\rho_{\rm pc}$.
	            
        \edit1{As noticed by W14, $\rho_{{\rm peak}}$ is probably greater than $|\beta|$ for pulsars with double-peak or multi-peak pulse profiles. The simplification of $\rho_{{\rm peak}} = |\beta|$ in Eq. ({\ref{eq:fan_Lum}}) will lead to overestimation of the luminosity for those pulsars, but it does not affect the results and conclusions essentially because of two reasons. Firstly, pulsars with double-peak and multi-peak profiles are less than 50\% \citep[e.g.][]{2007MNRAS.380.1678K}, while the remained pulsars with single-peak profiles are barely affected by this assumption. Secondly, even for double-peak or multi-peak profiles, the amount of overestimation is much smaller than the luminorsity dispersion that the dither factor describes. Considering an extreme case that the phase offset between the maximal pulse peak and the profile center reaches a half of the full pulse width, the phase offset is about $9^{\circ}$ when assuming a typical duty cycle of 5\%. Then the averaged $\Delta\rho/\rho_{\rm peak}$ is estimated to be $\sim$10\% for various inclination and impact angles, where $\Delta\rho=\rho_{\rm peak}-|\beta|$. Following Eq. ({\ref{eq:fan_Lum}}), the factor of lumonosity enhancement is $(1-\Delta\rho/\rho_{\rm peak})^{2q-6}\simeq1.4$ by taking $q=1.25$ (the optimized value obtained in this work), which is much smaller than the 1$\sigma$ dispersion of the luminosity induced by the dither factor, i.e. $10^{0.75}=5.6$. Therefore, the $\rho_{{\rm peak}} = |\beta|$ is a viable simplification.} 
	
	\subsection{Simulation Procedure and Common Models} \label{subsec:ProcModel}
	
        We describe the common models used in models A and B following the simulation procedure. These models and some related parameters, together with the models in the above subsections, are summarized in Table \ref{tab:EMA_pars}.  
	
        The simulation procedes as follows. First a synthetic pulsar is generated with a random age and a set of initial parameters. The age is chosen randomly between 0 and $t_{\rm max}$, with $t_{\rm max}$ being $10^9$ years\footnote{In principle $t_{\rm max }$ should be long enough to allow all the pulsars to cross the death line before reaching that age. It was found by FK06 that only a tiny fraction of synthetic pulsars are still alive, namely radio-loud after $10^9$ years, thus $t_{\rm max}=10^9$ years is practically a good choice.} (FK06, RL10). Its initial surface magnetic field $B$ is chosen from a log-normal distribution given by FK06, and its initial spin period $P_0$ is determined by a Gaussian distribution following FK06. The inclination angle $\chi$ is randomly selected by using Eq. (\ref{eq:chi}). 
	
        Next we let the pulsar evolve from $t=0$ to its age by using the spin-down model described in FK06, which only considers the magnetic dipole braking in a vacuum (see FK06 for details). Being the first trial of the fan beam model to be applied to pulsar population synthesis, this work is limited to a simple scenario in which the inclination angle and magnetic field do not evolve and the contribution of pulsar wind braking is ignored. This situation is equivalent to setting the braking index as 3. The current period is then calculated with 
	    \begin{equation}
	       P(t) = \sqrt{P_{0}^{2}+2kB^{2}t\sin^2{\chi} },
	    \end{equation}
        where $k=8\pi^{2}R^{6}/3Ic^{3}$, $I$ is the moment of inertia of the pulsar, assumed to be a typical value of $10^{38} \ {\rm kg \ m^{2}}$. Then the corresponding $\dot{P}$ can be computed by
	    \begin{equation}
    	    P\dot{P} = kB^{2}\sin{\chi}^{2}.
    	    \label{eq:ppdot}
	    \end{equation}
	
        The obtained current period is then compared with the death period, defined by the so-called death line relationship \citep{1992A&A...254..198B}
	   \begin{equation}
	       P_{{\rm death}} = \sqrt{\frac{B}{0.17 \times 10^{12} \ {\rm G \ s^{-2}}}},
	   \end{equation}
        beyond which the pulsar is regarded as being radio-quiet. Therefore the condition that a pulsar is radio-loud and potentially detectable is $P<P_{\rm death}$.

        Only radio-loud pulsars beaming towards the Earth are retained and procede to next steps. Then the spatial evolution of the synthetic pulsar needs to be modeled to determine its current position and distance, with which the corresponding dispersion measure (DM) can be determined in terms of a Galactic electron-density model. Next the model flux density, calculated from the luminosity and distance, and the observable pulse width affected by DM are used to estimate the theoretical S/N (cf. Section 4.1 in B14) to determine whether the pulsar is detectable by specific surveys or not.
	
	    The above procedure is basically the same as that in RL10 or B14. To model the initial position for each pulsar, here we choose the Gamma function described by L06 for the radial distribution, and the exponential distribution in FK06 for the distribution of Galactic height $z$. The birth velocity components of each pulsar in three coordinate directions are assigned by using the exponential distribution as per RL10. The method of modelling the spatial evolution of pulsars in the Galactic potential described in B14 is adopted here. We are aware that \citet{2017ApJ...835...29Y} has proposed a new version of electron-density model, but in this paper we still use the NE2001 eletron-density model incorporated in \textsf{\small{EVOLVE}}, in order to compare with the previous simulations with the conal beam model by FK06 and RL10. 
	
	    The selection of models in our model A is very similar to FK06 and RL10, except for the method to decide the initial Galactic coordinates of pulsars. FK06 or RL10 selected the radial distribution model proposed by \citet{2004AA...422..545Y} and used the spiral-arm stucture. We follow the method of model S in L06 by choosing a radial surface density model and an azimuthally symmetric function with respect to the Galactic centre. The default parameters of the models in the original \textsf{P{\scriptsize SR}P{\scriptsize OP}P{\scriptsize Y}} codes are used in this paper. In fact, we have tried some combinations of the radial and azimuthal models in our simulation, and found that the models we choosed can match the observation better, especially for the distribution of Galactic longitude.
	
	    Totally 1214 isolated pulsars (defined as $P > 30 {\rm ms}, 0 < \dot{P} < 1 \times 10^{-12}$) were detected in the PKSMB and PKSSW surveys, as cataloged in the ATNF pulsar catalogue V.1.62 \citep{2005AJ....129.1993M}. Each MC realization was run until 1214 pulsars were detected by these surveys in simulation (S/N $\geq$ 9). Following \citet{2004ApJ...604..775G}, we run 10 MC realizations for each simulation and compare our results with the observation to optimize the model parameters. 
	
        \startlongtable{}
        \begin{deluxetable}{ll}
    		\tablecaption{Models and parameters used in models A and B. \label{tab:EMA_pars}}
    		\startdata
    		Initial period distribution                    & Gaussian (FK06)\\
    		$\left \langle P_0\ {\rm (ms)} \right \rangle$ & 300 \\
    		std($P_0\ {\rm (ms)}$)                         & 150 \\
    		&\\
    		Initial $B$ field distribution         & Log-normal (FK06)\\
    		Model A                                & \\
    		$\left \langle {\rm log}_{10}B\ {\rm(G)}\right \rangle$    & 12.65 \\
    		std(${\rm log}_{10}B\ {\rm(G)}$)             & 0.55 \\
    		Model B   & \\
    		$\left \langle {\rm log}_{10}B\ {\rm(G)}\right \rangle$    & 12.25$^*$ \\
    		std(${\rm log}_{10}B\ {\rm(G)}$)             & 0.65$^*$ \\
    		&\\
    		Radial distribution model           & Gamma (L06) \\
    		Birth height distribution           & Exponential (FK06)  \\
    		$\left \langle {z_{0}\ {\rm(pc)}} \right \rangle$  	& 50 \\
    		Birth velocity distribution         & Exponential (FK06) \\
    		$\left \langle {v_{\rm l}\ {\rm (km}\ {\rm s}^{-1})}\right \rangle$    & 180 \\
    		&\\
    		Pulse width model                   & \\
    		Model A: conal beam                 & $W(\chi,\beta,\rho)$ relation (S09) \\ 
    		Model B: fan beam                   & $W(\chi,\beta,\varphi)$ relation (W14) \\
    		&\\
    		Beaming fraction ($f$) model        &  \\
    		Model A: conal beam                 & $f-P$ relation (TM98) \\ 
    		Model B: fan beam                   & $f=1^*$ \\
    		&\\
    		Inclination angle distribution ($\chi$)  & Random (S09) \\
    		Range                               & [0,90$^\circ$] \\
    		&\\
    		Impact angle distribution ($\beta$) & \\
    		Model A: conal beam                 & Uniform (S09)\\
    		Range                               & [$-\rho$, $\rho$] \\
    		Model B: fan beam                   & By $\beta=\zeta-\chi^*$\\
    		Range                               &  [$-90^\circ$,90$^\circ$] \\
    		&\\
    		Luminosity model                    & \\
    		Model A   & $L(P,\dot{P})$ function (FK06)  \\
    		Model B: fan beam  & $L(P,\dot{P}, \beta)$ function$^*$  \\
    		&\\ 
    		Spectral index distribution             & Gaussian \edit1{(B13)\tablenotemark{{\rm a}}} \  \\
    		$\left \langle \delta \right \rangle$   &  \edit1{-1.4} \ \\
    		std($\delta$)                           &  \edit1{0.96} \ \\
    		&\\
    		Eletron density model               & NE2001 (CL02)\tablenotemark{{\rm b}}\\
    		Scattering model                    & B04\tablenotemark{{\rm c}} \\
    		&\\
    		Pulsar spin-down model              & FK06 \\
    		Braking index                       & 3.0 \\
    		Max pulsar age (yr)                 & $10^{9}$  \\
    		&\\
    		Number of detectable pulsars in       & 1214 \\
    		the \edit1{PKSMB \& PKSSW}\,surveys &  \\
    		\enddata
    		\tablenotetext{}{*: This work. {\rm a}: \edit1{\citet{2013MNRAS.431.1352B}} ~\\ {\rm b}: \citet{2002astro.ph..7156C}, {\rm c}: \citet{2004ApJ...605..759B}}
    	\end{deluxetable}

	\section{Population Synthesis} \label{sec:Pop_Syn}
	
	\subsection{Assessment and Optimization}
\begin{figure*}[ht!]
    		\plotone{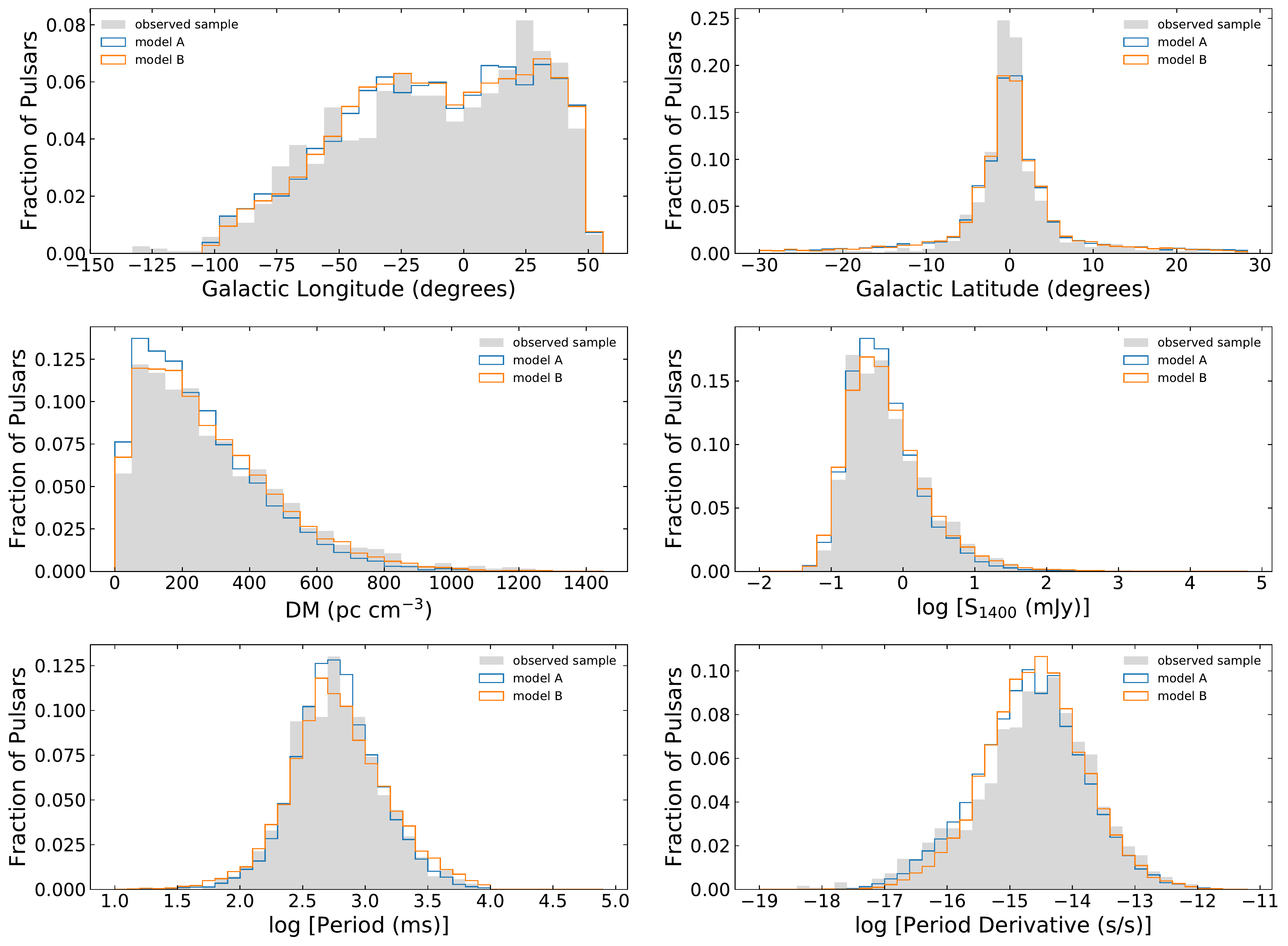}
    		\caption{Distributions of the Galactic longitude and latitude, dispersion measure (DM), 1.4-GHz \edit1{flux density} $S_{1400}$, spin period and period derivative for isolated radio pulsars. The grey histograms are for 1214 isolated pulsars discovered in the Parkes multibeam (PKSMB) and \edit1{Swinburne} pulsar surveys (PKSSW). The blue and red histograms stand for the isolated pulsars detected by the PKSMB and PKSSW surveys, which are simulated with the evolution model A (with conal beam) and B (with fan beam). For each evolution model, a total of $10\times1214$ pulsars from 10 MC realizations are used to derive the distributions.
    			\label{fig:pdf}}
    	\end{figure*}

        To assess the models, for each simulation, the observed marginal distributions of Galactic longitude ($l$), Galactic \edit1{latitude} ($b$), DM, flux density at 1.4 GHz ($S_{1400}$), $P$ and $\dot{P}$ in the simulation are compared with the corresponding distributions for the real sample, as shown in Fig. \ref{fig:pdf}. Our results with model A (blue) are very similar to those obtained by FK06 and RL10, despite some difference in model selection. It is shown that model B (orange) can reproduce the distributions of the real sample (gray) as well as model A, indicating that the fan beam is viable to simulate the population of isolated pulsars. Below we present the choices of parameters $B$, $\varphi$ and $q$ that produce the above results.
	
        Following FK06, a lognormal distribution of initial magnetic field with the mean value $\left \langle {\rm log}_{10}B\ {\rm(G)}\right \rangle=12.65$ and the standard deviation std$({\rm log}_{10}B\ {\rm(G)})=0.55$ is adopted in model A. Although it works in model A, it does not match the observations well when applied to model B, especially the distribution of $B$ and $\dot{P}$. A lognormal distribution with generally weaker field, i.e. with the mean value $\left \langle {\rm log}_{10}B\ {\rm(G)}\right \rangle=12.25$ and the standard deviation std$({\rm log}_{10}B\ {\rm(G)})=0.65$ is found to be able to reproduce the observed distributions with model B. 
	
        In the fan beam model of W14, $\varphi$ in Eq. (\ref{eq:fan_beam_width}) and $q$ in Eq. (\ref{eq:fan_I}) are two important parameters that affect the pulse width and luminosity, respectively. Their ``optimal'' values in model B are determined as follows. W14 constrained a range $[20^\circ, 80^\circ]$ for $\varphi$ by comparing the simulated {$W$} and $\beta$ with the real data of 64 pulsars. We tested different sets of $\varphi$ range close to $[20^\circ, 80^\circ]$ , including [0, 100], [20, 100] and [0,80], from which $\varphi$ is drawn from a uniform distribution. The simulations do not show obvious difference in the distributions of $l$, $b$, DM, $S_{1400}$, $P$ and $\dot{P}$. Since the simulation in W14 can well represent the boundary of the observed data in the \textbf{$W-\beta$} diagram (see Fig. 17 in W14), we still select the range $[20^\circ, 80^\circ]$ for $\varphi$.
    
        In W14, the value $q=1.75$ was obtained by fitting the pseudo luminosity at 1.4~GHz for a sample of 64 pulsars. Large dispersion in the luminosity data (cf. Fig. 18 in W14) leads us to suspect that this value may not be applicable to a much larger sample in the population simulation, hence it needs a procedure to perform optimization.

\begin{figure}[ht!]
    		\centering
    		\includegraphics[width=7cm,height=6cm]{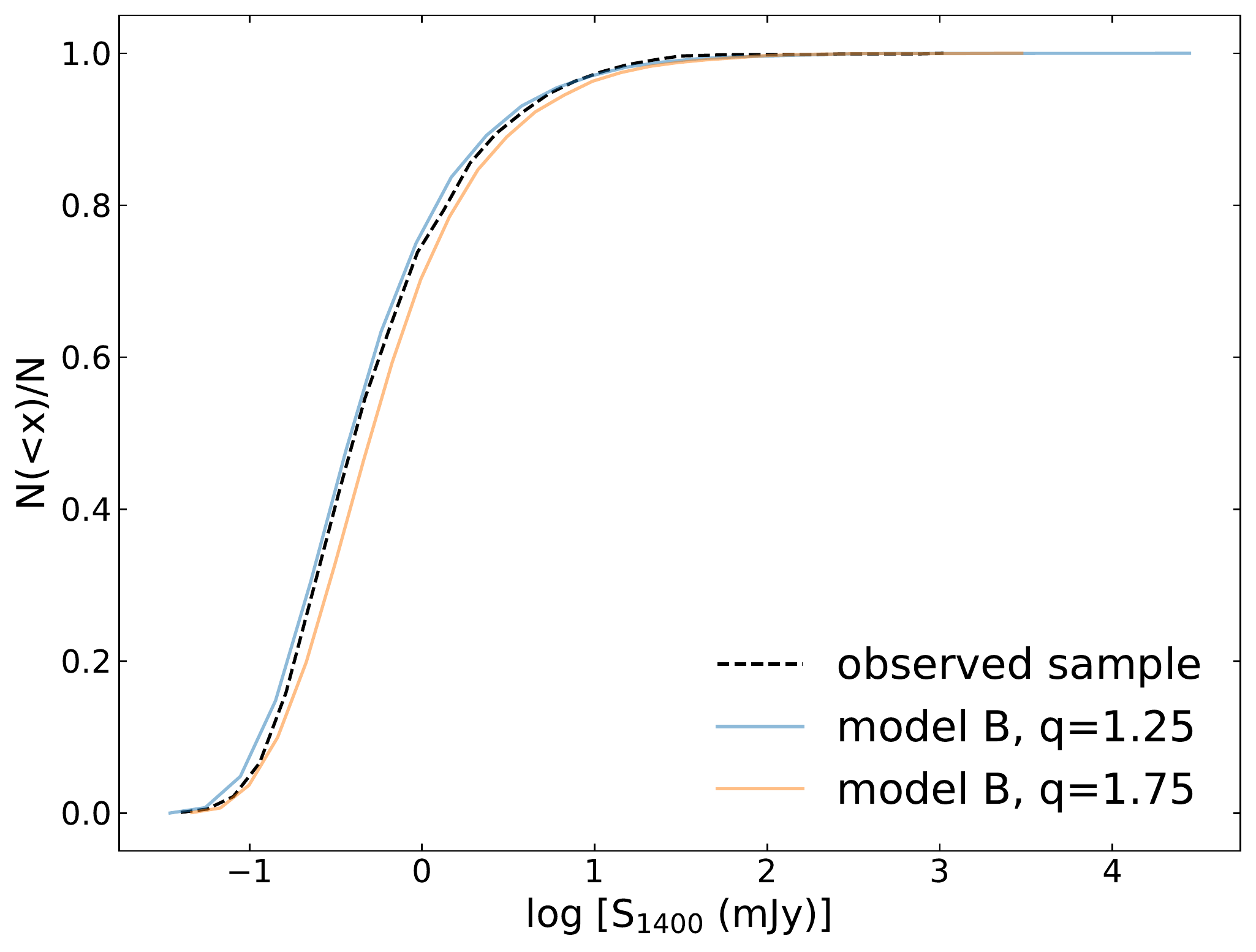}
    		\caption{Cumulative probability distributions of flux density at 1.4~GHz for the observed 1214 pulsars (dashed) and the simulated pulsars with model B in 10 MC realizations. The blue and orange curves are the results by using $q=1.25$ and $q=1.75$ in Eq (\ref{eq:fan_Lum}), respectively. The corresponding KS probability $Q_{S_{1400}}$ is \edit1{0.44} and \edit1{$9.15\times10^{-9}$}, suggesting that $q=1.25$ is a better solution than $q=1.75$. }
    		\label{fig:S1400_compare}
    	\end{figure}

\begin{figure}[ht!]
    		\centering
    		\includegraphics[width=7cm,height=6cm]{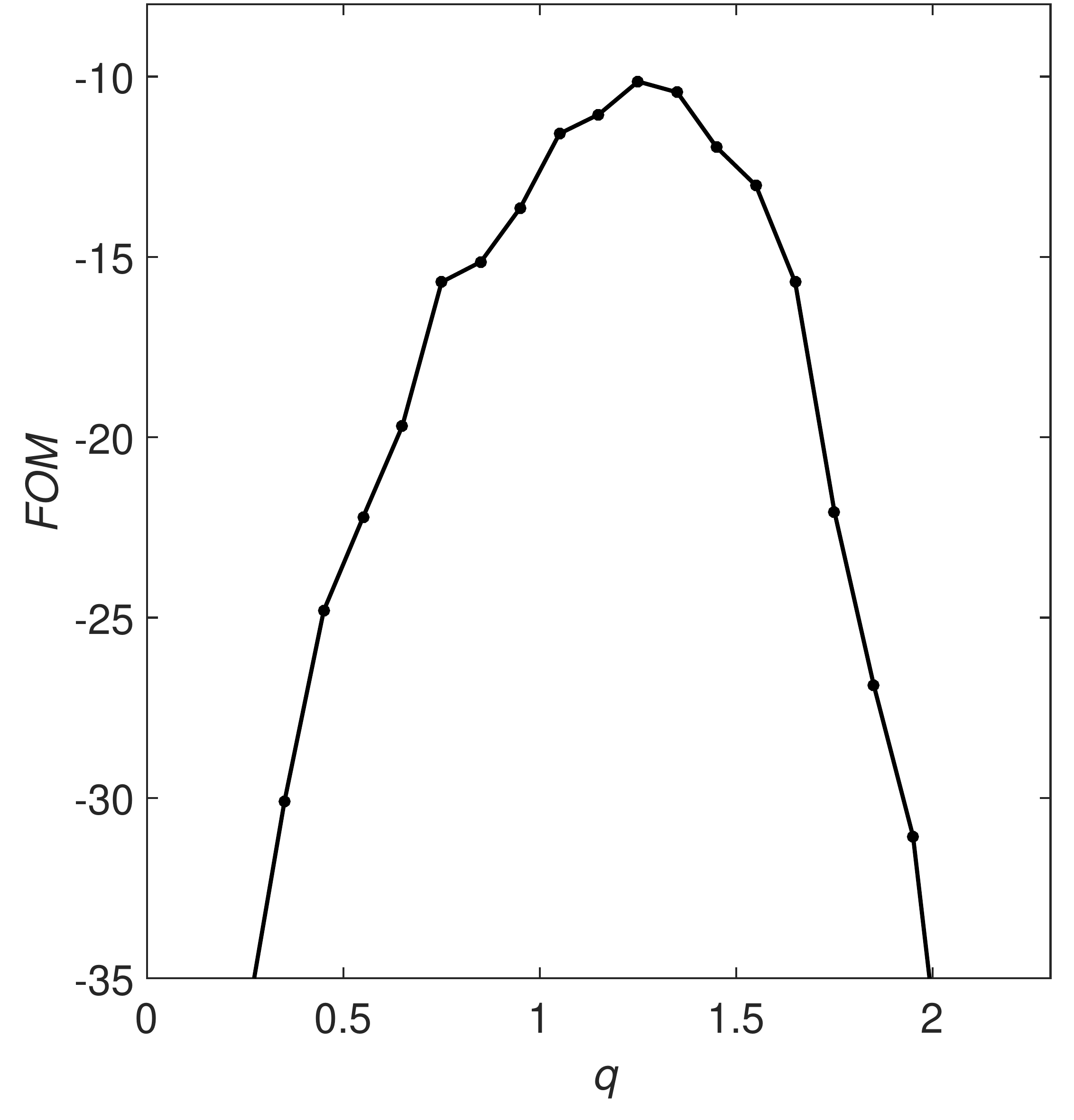}
    		\caption{The figure of merit (FOM) for each trial of $q$. For a given value of $q$, the FOM is calculated with Eq. (\ref{eq:fom}) to describe the relative goodness of modeling the observational distributions of isolated pulsars. In each simulation, all $10\times1214$ model pulsars from 10 MC realizations are used to calculate the FOM. The maximum FOM corresponds to $q$ = 1.25, which is used as the best parameter in the evolution model B.}
    		\label{fig:fom}
    	\end{figure}

	   Firstly we take $q=1.75$ for a trial. The main problem is that it slightly overpredicts the fraction of pulsars with large flux density compared with the $S_{1400}$ distribution of real sample (Fig. \ref{fig:S1400_compare}). In order to obtain the optimal value of $q$, we use the non-parametric Kolmogorov-Smirnoff (KS) test, and define a figure of merit (FOM) following \citet{1992A&A...254..198B} and RL10 for each simulation,
	   \begin{equation}
    	   {\rm FOM} = {\rm log}_{10}(Q_{l} \times Q_{b} \times Q_{\rm DM} \times Q_{S_{1400}} \times Q_{P} \times Q_{\dot{P}} ),
    	   \label{eq:fom}
	   \end{equation}
	   where the subscripts on each KS probability $Q$ correspond to the distributions of $l$, $b$, DM, $S_{1400}$, $P$ and $\dot{P}$, respectively. Obviously a larger FOM means a better simulation. To avoid the fluctuation from a single simulation, the associated KS probabilities are calculated with a total of $10\times1214$ model pulsars generated from 10 MC realizations. We then search the optimal $q$ value in a wide range of [0.25, 3.25] around 1.75. The resulted FOMs for given $q$ values are plotted in Fig. \ref{fig:fom}, in which the peak of the FOM curve indicates an optimal value $q=1.25$. This value is smaller than 1.75 suggested by W14, and the resulted distributions are consistent with those of the real sample (see Fig. \ref{fig:pdf}, also cf. Fig. \ref{fig:S1400_compare} to see the difference from the result of $q=1.75$).  
	
	\subsection{Underlying Populations of Isolated Pulsars}
	   Model A estimates nearly $1.81\pm0.04\times10^{5}$ \edit1{potentially observable pulsars that are radio-loud and  beaming towards the Earth}. The number is bit larger than $1.2\times10^{5}$ obtained by FK06. L06 inferred a total Galactic population of 30000$\pm$1100 pulsars with 1.4~GHz luminosities above 0.1 mJy~kpc$^2$, whereas our estimation is $\sim$ 85 000, more than 2 times of the L06's result. \edit1{Considering that fainter pulsars can be seen by telescopes with higher sensitivity, such as FAST, we also give an extra luminosity threshold of 0.01 mJy~kpc$^2$ to show the difference. Model A predicts about 156~000 pulsars whose luminosity are above 0.01 mJy~kpc$^2$.} 
	   
	   Large uncertainty exists in estimating the number of underlying radio-loud pulsars (including those not beaming towards us). Using a rough beaming fraction $f\simeq0.2$ for all periods (RL10), the total number of radio-loud pulsars is estimated to be $9.05\times10^{5}$, whereas when using the averaged fraction \textbf{$f\simeq0.1$} suggested by TM98, the total number would be $1.81\times10^{6}$. 
	   
	   Model B estimates about $2.27\pm0.07\times10^{6}$ potentially observable pulsars in the Galaxy, which is larger than the number of model A. This is partially because the fan beam is more extended than the conal beam, bringing more chance for \edit1{them} to be observed. In model B, \edit1{about 76~000 and 230~000 pulsars are estimated to have 1.4~GHz luminosities above 0.1 mJy~kpc$^2$ and 0.01 mJy~kpc$^2$, respectively.} 

       \begin{figure}[ht!]
            \centering
            \includegraphics[width=7cm,height=6cm]{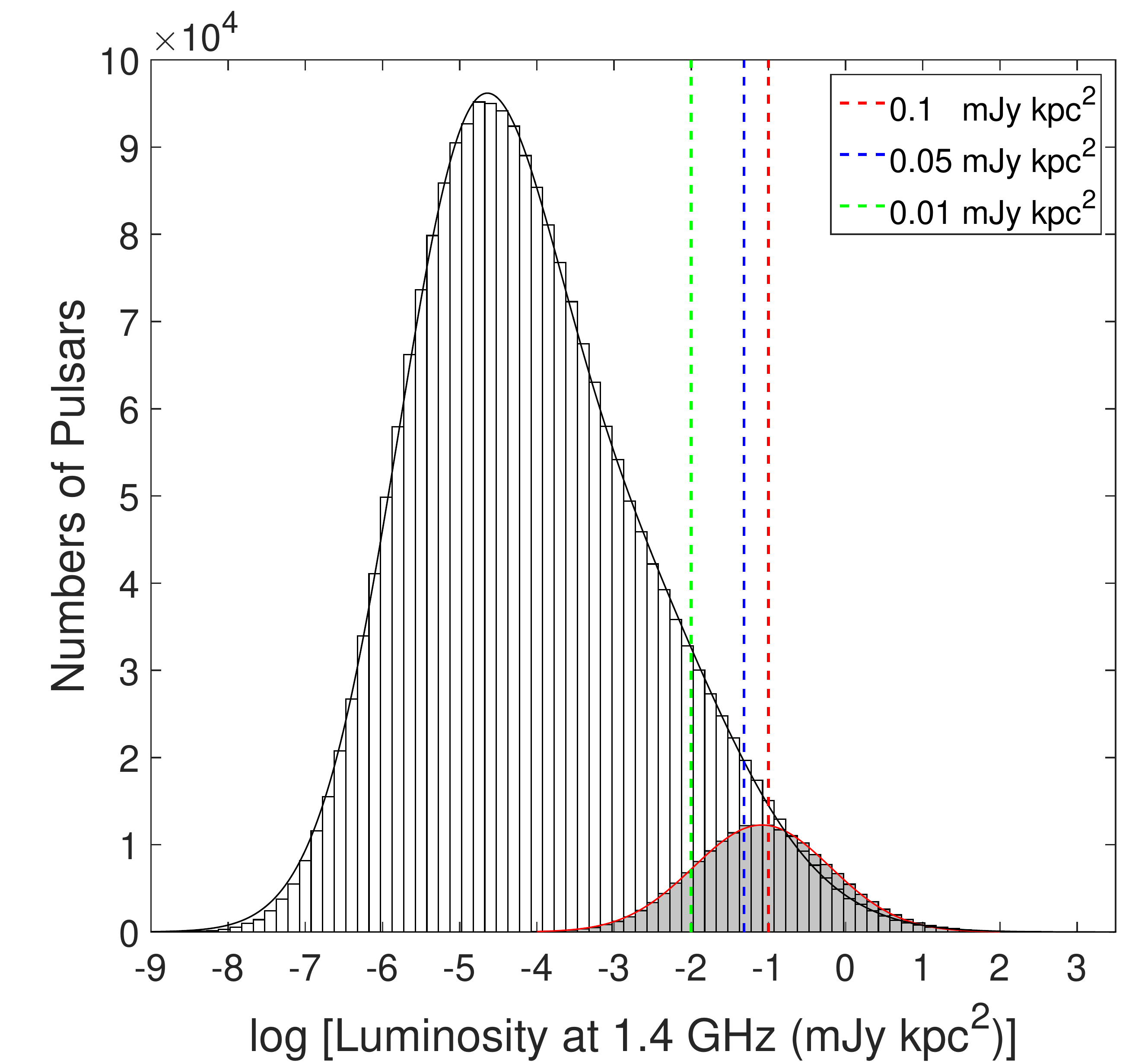}
            \caption{Distributions of 1.4-GHz pseudo luminosity for the underlying isolated pulsar population obtained with evolution models A (grey) and B (white), \edit2{respectively}. For each model, the sample includes the simulated pulsars that have not crossed the death line and are potentially observable (beaming towards us) in a MC realization. The red solid curve represents the best-fit lognormal function with mean value -1.1 and standard deviation 0.9, which is consistent with the result of FK06. The distribution of model B can be fitted by two lognormal functions, and the peak value of the best-fit curve corresponds to $L=2.3\times 10^{-5}$ mJy kpc$^{2}$. \edit1{The red and green dotted lines indicate the pseudo lumonosities of 0.1 mJy~kpc$^2$ and 0.01 mJy~kpc$^2$, respectively, while the blue dotted line represents the critical pseudo lumonosity (0.05 mJy~kpc$^2$), above which the numbers of pulsars of two models are equal.}}
            \label{fig:L_dist}
        \end{figure}
	
	   Fig. \ref{fig:L_dist} presents the distributions of 1.4~GHz pseudo luminosity for underlying radio-loud pulsars in a MC realization with models A and B, respectively. The peak value of pseudo luminosity of model B is about $2.3\times10^{-5}$ mJy ${\rm kpc^{2}}$, which is about 3 orders of magnitude lower than that of model A ($7.9\times10^{-2}$ mJy ${\rm kpc^{2}}$). Our result of model A is generally consistent with FK06, in which the underlying distribution of luminosity follows a lognormal distribution with mean value -1.1 ($L={\rm 0.07 \ {\rm mJy} \ {\rm kpc}^{2}}$). The weak  predominance of model-A pulsars with pseudo lunimosity above 0.1 mJy~kpc$^2$ can also be seen from the plot.

\begin{figure}[ht!]
    		\centering
    		\includegraphics[width=7cm,height=6cm]{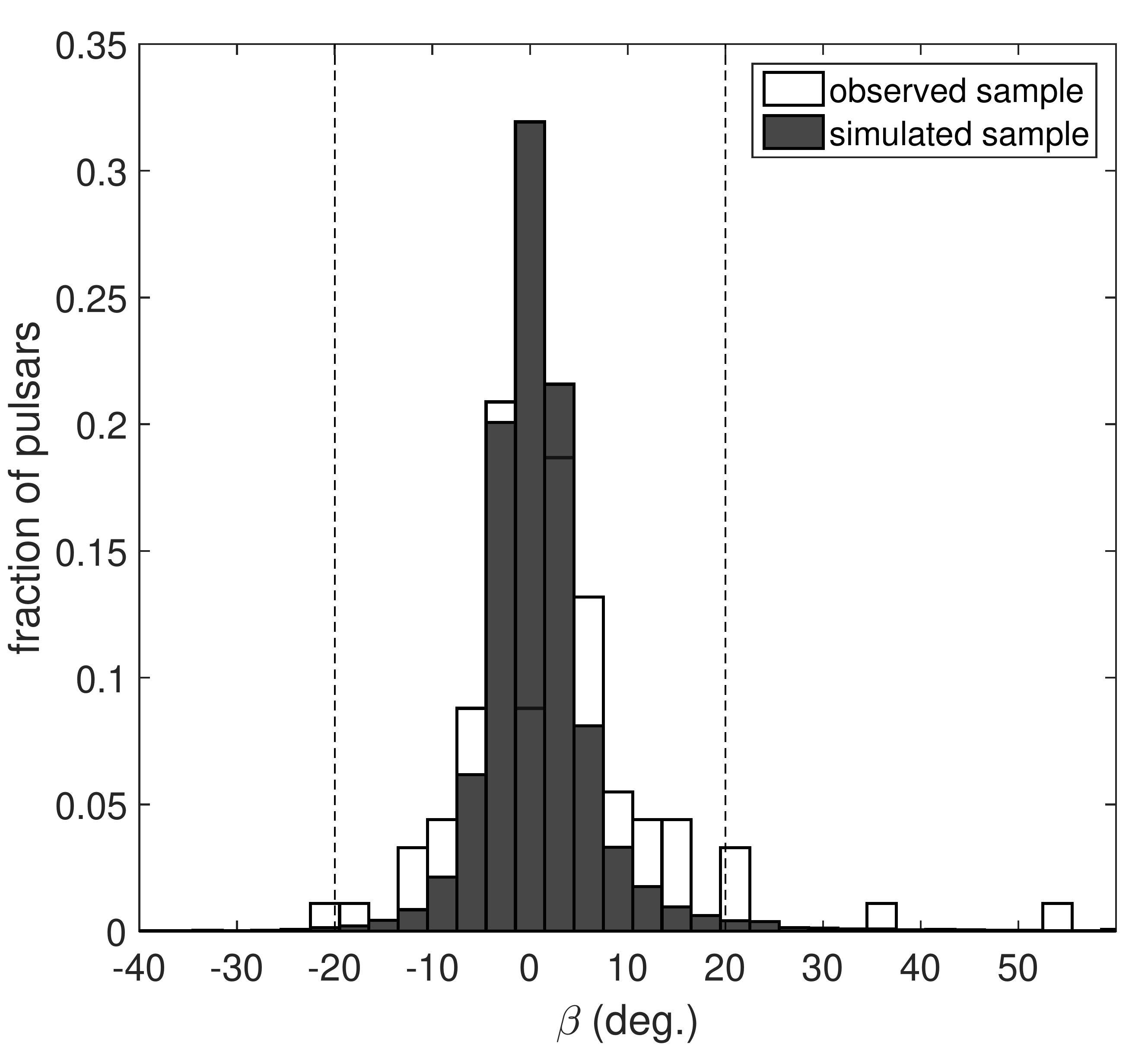}
    		\caption{The distribution of impact angle $\beta$ of model pulsars detected by the surveys of PKSMB and PKSSW (dark) and the distributon of the observed $\beta$ for a sample of 79 pulsars. The observational sample consists of 60 pulsars (12 of them have the main pulse and the interpulse) from W14 and 19 pulsars from \citet{2015MNRAS.446.3367R}. The simulation is roughly consistent with the observed distribution. Most of the impact angles of simulation pulsars are smaller than 20$^{\circ}$.}
    		\label{fig:beta_dist}
    	\end{figure}

	   The abundance of weaker pulsars in the population of model B is expected by the fan beam model. This is becuase the fan beam model tends to predict a large amount of pulsars with large impact angles, and their pseudo luminosity \edit1{is very low} because of the radial limb-darkening relationship (Eq. \ref{eq:fan_Lum}). Such weak pulsars with large impact angle are hard to detect. To illustrate this point, we plot the histogram of \edit1{impact} angle for the simulated pulsars observed with the PKSMB and PKSSW surveys (Fig. \ref{fig:beta_dist}). It shows that most detected pulsars has small impact angles with $|\beta|<20^\circ$, and the simulated result is generally consistent with the observed data of $\beta$, which are taken from W14 and \citet{2015MNRAS.446.3367R}\footnote{The sample of data includes 72 $\beta$ values of 60 pulsars from W14 and 19 values from \citet{2015MNRAS.446.3367R}.  Among the pulsars from W14, 12 pulsars have values for both the main pulse and the interpulse. In \citet{2015MNRAS.446.3367R}, 25 pulsars have the constrained $\beta$ values, but 4 pulsars with errors larger than $|\beta|$ and 2 pulsars overlapped with W14 are discarded.}.

	\subsection{$P$-$\dot{P}$ Diagram}      
	
\begin{figure*}[ht!]
    		\plotone{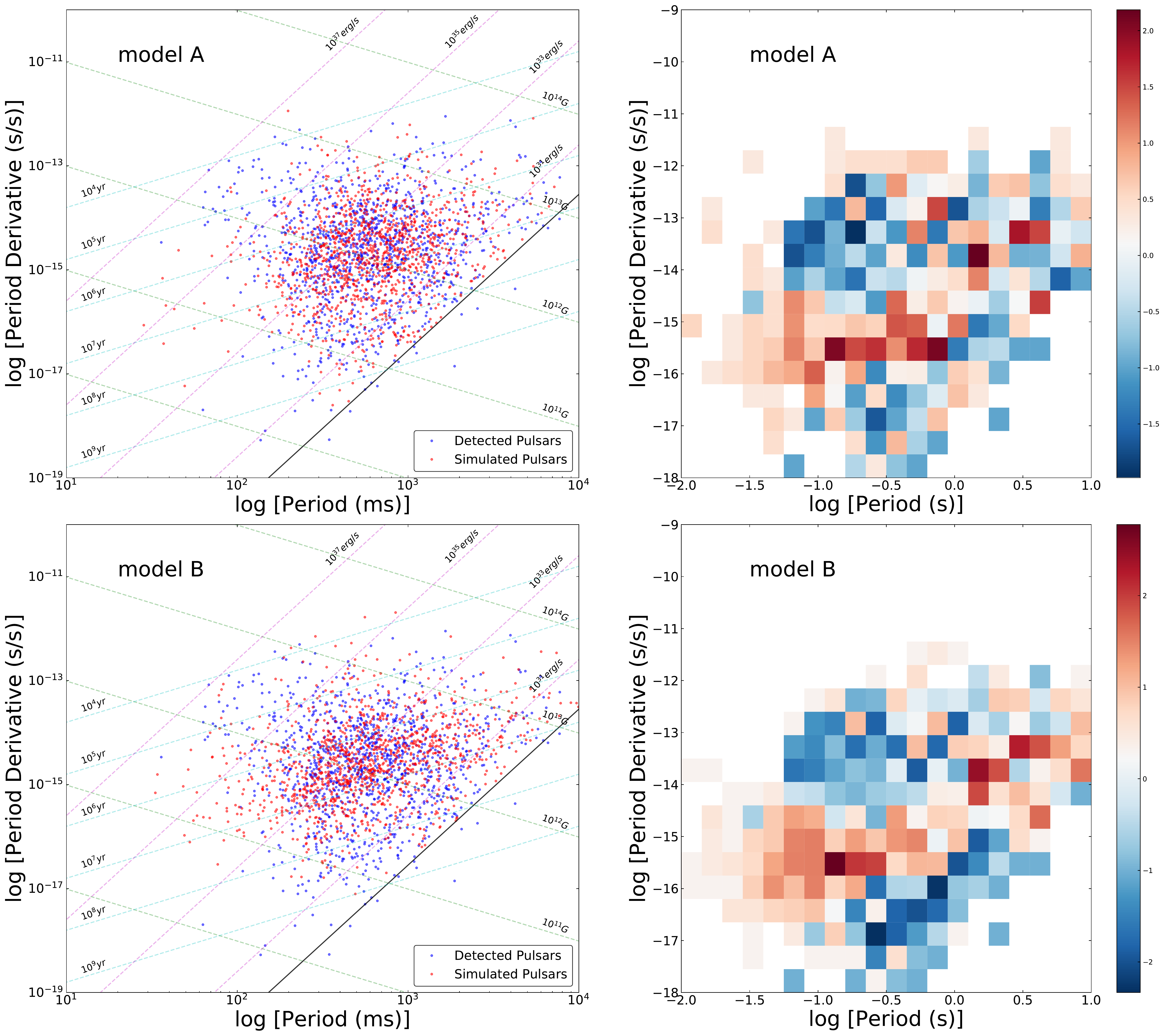}
    		\caption{Left: $P$-$\dot{P}$ diagrams for the actual pulsars (blue) and simulated pulsars (red) detected by the PMSMB and PKSSW surveys. Contours of surface magnetic field, characteristic age and loss rate of rotational kinetic energy are plotted by dashed lines in different colors, respectively. The thick solid line represents the theoretical death line. Right: the difference between the numbers of the observed and simulated pulsars in 2-D $P$-$\dot{P}$ bins. Red patches indicate $\edit1{\overline{N}_{{\rm sim}}}>N_{{\rm obs}}$, while blue patches indicate $\edit1{\overline{N}_{{\rm sim}}}<N_{{\rm obs}}$. See Eq. (\ref{eq:R_ppdot}) for details.
    		\label{fig:PPdot}}
    	\end{figure*}

	   The marginal distribution of $P$ and $\dot{P}$ can not reflect enough details about the difference between the simulated and real samples. Diagnosis using the $P$-$\dot{P}$ diagram is helpful to detailed studies. The real pulsars and the simulated pulsars detected in PMSMB and PKSSW surveys for a MC simulation are plotted in Fig. \ref{fig:PPdot} for \edit1{models A and B}. We follow \cite{2017MNRAS.467.3493J} to anaylize the plot quantitatively by calculating the difference factor
    	\begin{equation}
        	R = \frac{\edit1{\overline{N}_{{\rm sim}}}-N_{{\rm obs}}}{\sqrt{\edit1{\overline{N}_{{\rm sim}}}+N_{{\rm obs}}}},
        	\label{eq:R_ppdot}
    	\end{equation}
	   for each two-dimensional (2D) bins in the $P$-$\dot{P}$ plane, \edit1{where $N_{{\rm obs}}$ is the number of the observed pulsars and $\overline{N}_{{\rm sim}}$ is the averaged number of the model pulsars in 10 MC simulations, respectively.} The color scale is set by the value of $R$, with red for positive values (simulated pulsars are more than real ones) and blue for negative values (simulated pulsars are less than real ones). One can see that \edit1{both model A and model B do} not work very well in the top left-hand side and the bottom right-hand side in the diagram, \edit1{while model B predicts less pulsars than model A does in the bottom right-hand side. This can be explained by the difference in the relationships between the luminosity and the $P$ and $\dot{P}$ parameters. In model A, following Eq. (\ref{eq:FK06_Lum}) and ignoring the dither factor, luminosity contours in the $P$-$\dot{P}$ diagram are lines with a slope rate of 3, which are parallel to $\dot{E}$ contours. In model B, as the luminosity of fan beam is proportional to $P^{-2.75}\dot{P}|\beta|^{-3.75}$, the luminosity contour lines have a smaller slope rate of 2.75 (ignoring the dither factor and the dispersion caused by the $\beta$ term). The difference in the luminosity relationships is passed through the simulations and hence affects the distributions of simulated samples of two models.}
       
       \edit1{It is also noticed that both models overpredict the number of pulsars with P $\leq$ 100 ms and $\dot{P}$ between $10^{-16}$~s/s and $10^{-15}$~s/s. This is because under the distributions of initial spin period and magnetic field and the simple dipole magnetic braking model that we used, a considerable fraction of pulsars can evolve into that range of $P$ and $\dot{P}$. Those pulsars spin fastly and have moderate $\dot{P}$ values, therefore may be bright enough to be detected by the surveys. To overcome this problem, more comprehensive spin-down models involving the evolution of magnetic field or inclination angle and alternative choices of initial parameters need to be employed \citep[e.g.][]{2017MNRAS.467.3493J}.} This will be studied in a futher work.
	
	\section{Application in FAST Detection}\label{sec:FAST_dectection}

	   The above optimal population of evolution models A and B can be used to predict how many pulsars that future pulsar surveys may detect, through the \edit1{\textsf{\small{DOSURVEY}} module} in \textsf{P{\scriptsize SR}P{\scriptsize OP}P{\scriptsize Y}}. In this paper we show its application to FAST pulsar survey. S09 has performed simulations based on ``snapshot model'' to estimate the number of pulsars that FAST can detect. Here, following the method of S09 but with some updated parameters, we also perform snapshot simulations for FAST surveys by using the \edit1{\textsf{\small{POPULATE}} module} in \textsf{P{\scriptsize SR}P{\scriptsize OP}P{\scriptsize Y}}. Then we compare the simulations results of the snapshot model and the evolution models A and B. 
	
	   Note that the difference of the snapshot models by S09 and this paper, is that we still adopt the method of model S in L06 to determine the Galactic coordinates of pulsars, while S09 used the model C$^\prime$ in L06, following the spiral-arm modeling procedure described by FK06. Except that, other models (e.g. the distribution of $P$, $z$, $L$) are basically consistent. Table \ref{tab:fast_pars} lists the parameters of FAST multibeam survey used in the simulation. The frequency range is from 1050~MHz to 1450~MHz, with the center frequency 1250 MHz and the bandwidth 400 MHz \citep{2019SCPMA..6259502J}. This is slightly different from S09, who used the same bandwith but a differnet centre frequency of 1315~MHz. The same as above, we generate the population with the snapshot model until 1214 pulsars are detected by simulations of PKSMB and PKSSW surveys. Finally, we obtained the underlying sample of 125 000$\pm$4 000 model pulsars, which is close to the results of S09 with 120 000 pulsars.
	
    	\startlongtable
    	\begin{deluxetable}{lc}
    		\tablecaption{The survey parameters of FAST \label{tab:fast_pars}}
    		\tablehead{ \colhead{Parameters} & \colhead{Values} }
    		\startdata
    		Antenna gain (K/Jy)                    & 16.5   \\
    		Sampling time (ms)                     & 0.05   \\
    		System temperature (K)                 & 25     \\    
    		Centre frequency (MHz)                 & 1250   \\    
    		Bandwidth (MHz)                        & 400    \\
    		Channel bandwidth (MHz)                & 0.39   \\    
    		Number of polarizations                & 2      \\
    		Full-width half maximum (arcmin)       & 2.94   \\       
    		Minimum DEC (deg.)                     & -14  \\    
    		Maximum DEC (deg.)                     & 66   \\ 
    		\enddata
    	\end{deluxetable}
	
\begin{figure*}[ht!]
    		\centering
    		\plotone{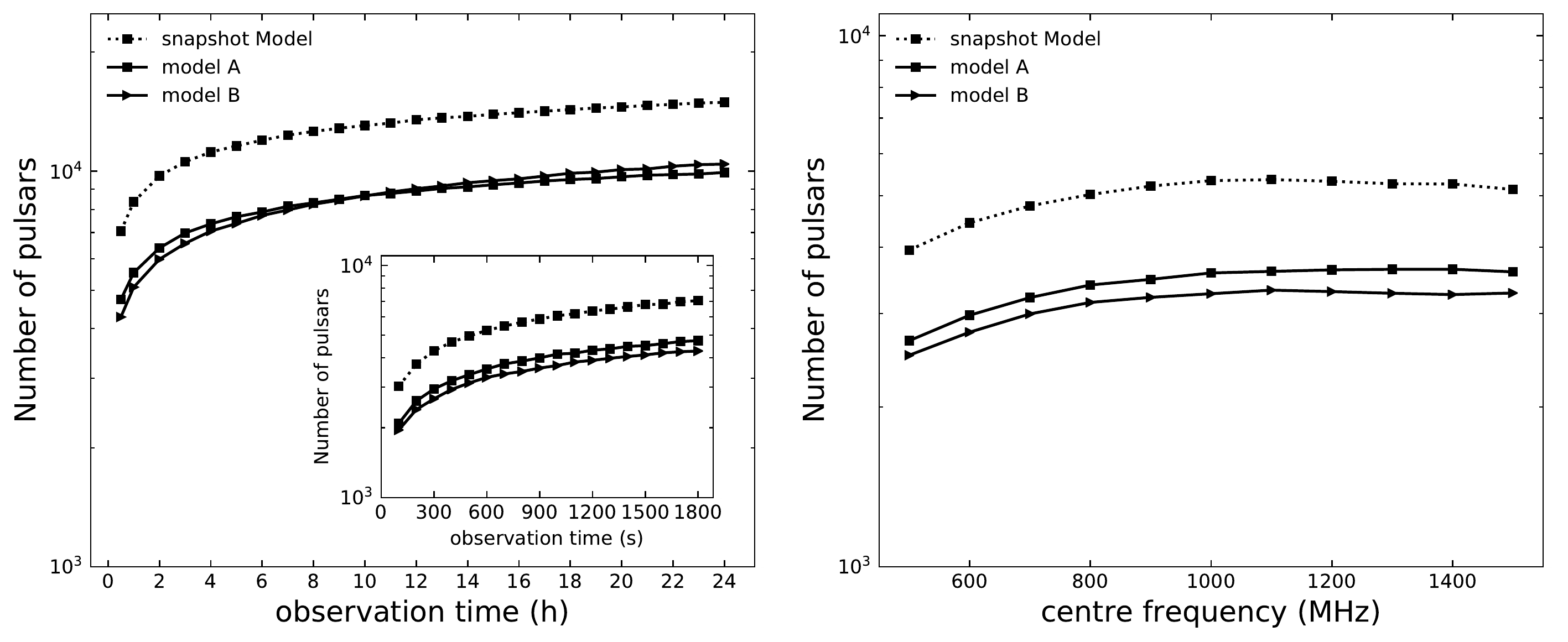}
    		\caption{Number of isolated pulsars detected by simulated FAST survey as a function of observation time per \edit2{pointing} (left) and the centre frequency (right). Different types of broken lines represent the results of three models, respectively. The regions of surveys are limited in galactic plane within $|b| < 10^{\circ}$ and $20^{\circ} < l < 90^{\circ}$. \edit1{In the left panel, the inset presents the results for observations within half an hour.} In the right panel, the bandwidth is set to one-third of each centre frequency that ranges from 500 MHz to 1500 MHz, and the observation time per pointing is unified as 600s. Note that here the detection threshold of signal-to-noise ratio (S/N) is 9, the same as in S09.}
    		\label{fig:time_freq_No}
    	\end{figure*}

        Figure \ref{fig:time_freq_No} shows the number of isolated pulsars detected in simulations of FAST surveys of inner Galactic plane (defined as $20^\circ < l < 90^\circ$ and $|b|<10^\circ$) as a function of observation time per pointing (left panel) and a function of observation frequency (right panel) for the snapshot model, models A and B, respectively. In the right panel, the increasing trends of the curves are very similar, showing that the rate of increase gets slower as the observation frequency increases, and the number becomes stable in the frequency range between 1 GHz and 1.5 GHz. 
        
        \edit1{In the left panel, it can be seen that model B predicts fewer detectable pulsars than model A does in short observation time (see the inset). However, when the observation time increases, the number of detected pulsars by model B gradually approaches the number of pulsars by model A, and surpasses it when the observation time is longer than 11 hours (with the sensitivity higher than $\sim 1\mu$Jy with a signal to noise ratio of 9). This is because the number of pulsars of model A slightly predominates over that of model B in high pseudo luminosities, while it is reversed in low \edit2{luminosities}. Therefore, model B predicts discoveries of a large number of low-luminosity pulsars for highly sensitive surveys.} 
       
        Both of the evolution models predict much less detectable pulsars than the snapshot model. For example, using an observation time 600~s per pointing, the expected numbers of pulsars to be detected are about 5300, 3600 and 3300 for the snapshot model, models A and B, respectively. Substracting the known \edit1{$\sim$ 600} isolated pulsars in the region, about \edit1{4700}, \edit1{3000} and \edit1{2700} new isolated pulsars are expected to be discovered by the FAST survey. Note that the result of our snapshot model is a bit smaller than that predicted by S09. When using the system parameters of FAST survey listed in S09, the expected number increases to 5600, but is still smaller than the number 6300 obtained by S09. This may be caused by the different choices of models or pulsar surveys in our MC simulations. Considering that real pulsars should evolve, we suggest that the snapshot model may overpredict the yields of FAST survey. 
		
        If the FAST Galactic-plane survey is carried out in a different longitude range $30^{\circ} < l < 100^{\circ}$, the number will decrease to about 4000, 2800 and 2600 for the above three models, respectively. Correspondingly, the unkown pulsars will be about \edit1{3600}, \edit1{2400} and \edit1{2200}, respectively. For the outer Galactic plane (defined as $150^{\circ} < l < 210^{\circ}$, $|b| < 10^{\circ}$), model B predicts around 290 pulsars expected to be detected with FAST, including \edit1{$\sim$ 50} known pulsars.

\begin{figure*}[ht!]
    		\centering
    		\includegraphics[width=10cm,height=5.7cm]{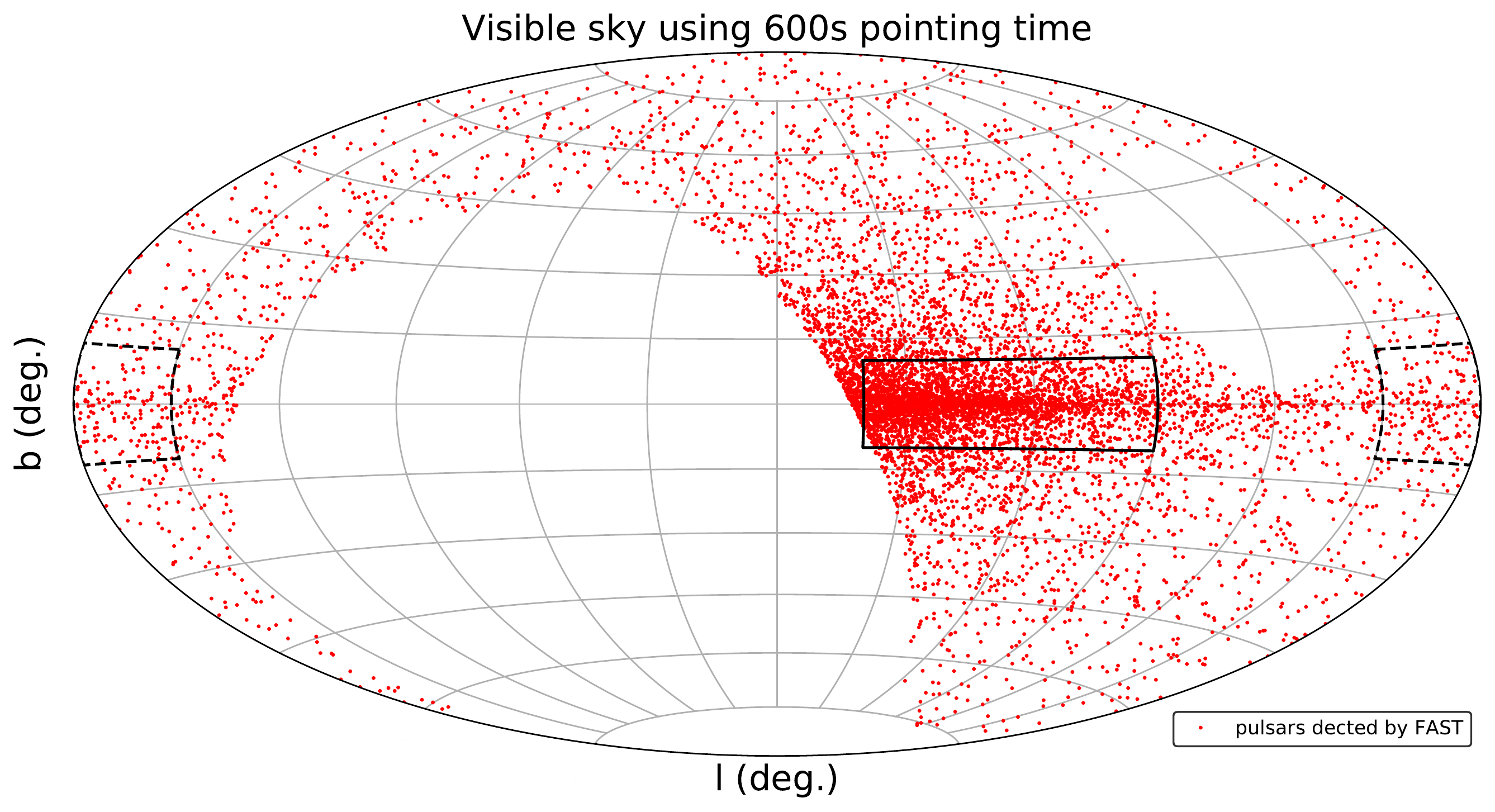}
    		\caption{A Hammer-Aitoff projection of the distribution of isolated pulsars (red dots) detected by FAST from model B in galactic coordinates. The black solid box marks the region in $20^{\circ} < l < 90^{\circ}$, $|b| < 10^{\circ}$, while the black dotted boxes represent the region $150^{\circ} < l < 210^{\circ}$, $|b| < 10^{\circ}$. See text for details.}
    		\label{fig:fast_proj}
    	\end{figure*}

        Fig. \ref{fig:fast_proj} shows the distribution of isolated pulsars detected by FAST in the Galaxy simulated with model B.  With 600 s integration time, totally 7700 pulsars could be detected (including pulsars already known) in the field of view of FAST. The inner and outer Galactic planes that are visible to FAST are marked by the solid and dotted black boxes, respectively. 
	
	\section{Conclusion and discussion}\label{sec:conclusions}
	
    	In previous pulsar population synthesis incorporating emission beam model, the radio beam is usually assumed to be circular, and the radio luminosity is regarded to be independent of beam geometry. In contrast to the conal beam model, the fan beam model proposed by W14 predicts the opposite trend of $W$-$|\beta|$ relationship and an viewing-geometry-dependent luminosity function. In this paper, we applied the fan beam model to perform population synthesis for isolated radio pulsars. 
        
        On the basis of \textsf{P{\scriptsize SR}P{\scriptsize OP}P{\scriptsize Y}} software package, we developed a code incorporating the pulse width and luminosity relationships of the fan beam model. Adopting a set of models from literature for the initial spin period and magnetic field, birth location and velocity, Galactic potential etc, and assuming that puslars spin down via dipole magnetic braking, we performed simulations for two evolution models incorporating the conal beam and fan beam models, respectively. By comparing the simulated distributions of $l$, $b$, DM, $S_{1400}$, $P$ and $\dot{P}$ with the observed distributions of the real 1214 isolated pulsars decoveried by the PKSMB and PKSSW surveys at 1.4~GHz, we found that the evolution model with fan beam can reproduce the observed distributions as well as the conal beam model can do. Major results are as follows.
        
        (1) Using the population synthesis, the optimal luminosity function at 1.4~GHz of the fan beam model is proportional to $P^{q-4}\dot{P}_{-15} |\beta|^{2q-6}$ with $q=1.25$. This result updates the former value $q=1.75$ obtained by W14 from a small sample of pulsars.  
        
        (2) The evolution model with fan beam predicts a Galactic population of about $2.27\times10^{6}$ isolated pulsars potentially observable, among which \edit1{$2.30\times10^{5}$} pulsars have pseudo luminosities at 1.4~GHz above \edit1{0.01} mJy~kpc$^2$. While these numbers are $1.81\times10^{5}$ and \edit1{$1.56\times10^{5}$} for the evolution model with conal beam, respectively. Assuming a beaming fraction of 0.2, the radio-loud isolated pulsars is estimated to be $9.05\times10^{5}$ in the case of conal beam model.   
        
        (3) The population synthesis incorporating the fan beam model is applied to predict the yields of FAST pulsar survey. For purpose of comparison, we also performed simulations with the snapshot method similar to S09. For an inner Galactic-plane pulsar survey in the region $20^\circ < l < 90^\circ$ and $|b|<10^\circ$ and with an integration time of 10 minutes per pointing at 1.25~GHz, the evolution model with fan beam predicts that \edit1{2700} unkown pulsars may be discovered, which is close to the number of \edit1{3000} predicted by the evolution model with conal beam. However, the snapshot model predicts a much more yield of \edit1{4700} new pulsars, which may be over optimistic. For the outer Galactic plane ($150^\circ < l < 210^\circ$ and $|b|<10^\circ$), the number of unknown pulsars detected by FAST surveys is expected to be \edit1{240} by the evolution model with fan beam. 
        
        It should be noted that the luminosity relationship with $q=1.25$ used in the evolution model with fan beam is obtained by searching for the parameter space of $q$, whereas the coefficient $\kappa$ is fixed as the value obtained by W14. In general, for a given value of $\kappa$, the number of underlying pulsars increases with decreasing $q$. This is because smaller $q$ values will lead to faster intensity atenuation with increasing $|\beta|$, hence reducing the detection probability for synthetic pulsars. A larger number of underlying pulsars is then needed to balance this effect. In this paper, \edit1{we} did not perform the computational demanding optimization for both $\kappa$ and $q$, leaving it elsewhere to be explored. The resultant luminosity relationship should be regarded as a viable solution to match the observations.
        
        It is still an open question whether the radio fan beam has an abrupt radial boundary. The precessional pulsar PSR J1906+0746 is a unique pulsar of which the main pulse and interpulse radio emission beams are in the shape of fan beam. It was found that the main-pulse emission was no longer detected in individual observations with 305-m Arecibo radio telescope when $\edit1{|\beta_{{\rm MP}}|}>22^{\circ}$ since MJD \edit1{57713} \citep{2019Sci...365.1013D}. However, using the online data\footnote{https://doi.org/10.5281/zenodo.3358819} between MJD \edit1{57713} and MJD 58290, we found a weak main pulse when integrating those $\sim$ 20-hour data. The flux is estimated to be $\sim2\mu$Jy. This indicates that the fan beam actually extends to a further radial distance. More sensitive observations in the future, e.g. with FAST are helpful to trace the extent of the radio beams of this pulsar. In this paper, we simply assume that the fan beam can extend to 90$^\circ$ away from the magnetic pole, and found the observations can be reproduced well under this assumption. If there is really an \edit1{abrupt} boundary making the extent less than 90$^\circ$, the underlying population of isolated radio-loud pulsars would be smaller than that estimated in Section \ref{sec:Pop_Syn}.
        
    \acknowledgments
        We appreciate Chen Wang for helpful discussion. This work is supported by National Natural Science Foundation of China No.11573008. 
        We acknolowdge supports from the Astronomical Science and Technology Research Laboratory of Department of Education of Guangdong Province and the Key Laboratory for Astronomical Observation and Technology of Guangzhou.
        H. G. Wang is supported by the 2018 Project of Xinjiang Uygur Autonomous Region of China for Flexibly Fetching in Upscale Talents. 
        We would like to thank Bates et al for sharing the \textsf{P{\scriptsize SR}P{\scriptsize OP}P{\scriptsize Y}} software package in the Github. We also made use of the ATNF pulsar catalogue at http://www.atnf.csiro.au/people/pulsar/psrcat/ for this work. 

        ~\\\edit1{\software{\textsf{P{\scriptsize SR}P{\scriptsize OP}P{\scriptsize Y}} \citep{2014MNRAS.439.2893B}}}
	
	\bibliography{References}{}
	\bibliographystyle{aasjournal}

\end{document}